\documentclass[a4paper,11pt]{article}
\pdfoutput=1 

\usepackage{jheppub} 
\usepackage{subcaption}
\newcommand{\orcid}[1]{\href{https://orcid.org/#1}{\includesvg[width=10pt]{orcid}}}

\counterwithout{equation}{section}
\usepackage[T1]{fontenc} 

\usepackage{url}
\usepackage{feynmf}
\usepackage{braket}
\usepackage[capitalise]{cleveref}
\usepackage{ulem}

\title{\boldmath On neutrino-mediated potentials in a neutrino background}
\author[a,b]{Diego Blas}
\author[c,d]{Ivan Esteban}
\author[e,f,g]{M.C. Gonzalez-Garcia}
\author[e]{Jordi Salvado}

 \affiliation[a]{Grup de Física Te\`orica, Departament de Física, Universitat Aut\`onoma de Barcelona,\\ 08193 Bellaterra (Barcelona), Spain}
 \affiliation[b]{Institut de Física d’Altes Energies (IFAE), The Barcelona Institute of Science and Technology, Campus UAB, 08193 Bellaterra (Barcelona), Spain}
\affiliation[c]{Center for Cosmology and AstroParticle Physics (CCAPP), Ohio State University, 191 W Woodruff Ave., Columbus, OH 43210}
\affiliation[d]{Department of Physics, Ohio State University,  191 W Woodruff Ave., Columbus, OH 43210}
\affiliation[e]{Departament de Física Quàntica i Astrofísica and
  Institut de Ciències del Cosmos, Universitat de Barcelona, Diagonal
  647, E-08028 Barcelona, Spain}
\affiliation[f]{Institució Catalana de Recerca i Estudis Avançats (ICREA), Barcelona, Spain} 
\affiliation[f]{C.N. Yang Institute for Theoretical Physics, Stony Brook University, Stony Brook, USA}

\emailAdd{dblas@ifae.es}
\emailAdd{Maria.Gonzalez-Garcia@stonybrook.edu}
\emailAdd{esteban.6@osu.edu}
\emailAdd{jsalvado@icc.ub.edu}

\abstract{The exchange of a pair of neutrinos with Standard Model weak
  interactions generates a long-range force between fermions. The
  associated potential is extremely feeble, $\propto G_F^2/r^5$ for massless
  neutrinos, which
  renders it far from observable even in the most sensitive
  experiments testing fifth forces. The presence of a neutrino
  background has been argued to induce a correction to the neutrino
  propagator that enhances the potential by orders of magnitude.  In
  this brief note, we point out that such modified propagators are
  invalid if the background neutrino wavepackets have a finite width.
  By reevaluating the 2--$\nu$ exchange potential in the presence of a
  neutrino background including finite width effects, we find that the
  background-induced enhancement is reduced by several orders of
  magnitude. Unfortunately, this pushes
  the resulting 2--$\nu$ exchange potential away from present and
  near-future sensitivity of tests of new long-range forces.}

\begin{document}
\normalem
\maketitle
\flushbottom

\section{Introduction: short review of neutrino-mediated forces}

Soon after the existence of neutrinos was proposed, it was pointed out
that the exchange of neutrino pairs would induce a long-range force
between nucleons~\cite{Bethe:1936zz,gamow-teller,rusos}.  Most of
these early works estimated the potential to behave as $1/r$, with $r$
the distance between two test particles, and even suggested that it
could be responsible for
gravitation~\cite{gamow-teller,gamow,corben,bocchieri}.  However, as
early as 1936, Iwanenko and Sokolow~\cite{rusos} obtained that the
potential due to the exchange of two massless fermions behaves as
$1/r^5$.  This did not prevent Feynman to consider a $1/r^3$
dependence in Ref.~\cite{Feynman:1963ax, Feynman:1996kb}, reinforcing
the idea that neutrinos could be mediators of a long-range interaction
sourced by mass.

The confusion in the literature concerning the $r$-dependence of the
potential was finally resolved by Feinberg and
Sucher~\cite{Feinberg:1968zz}, who carried out the first calculation
of the potential using the effective four-fermion interaction of weak
charged currents of the Standard Model (SM) via dispersion techniques
(the calculation was later complemented to include the neutral current
contribution~\cite{Feinberg:1989ps}).  They found that, for the
exchange of massless neutrino-antineutrino pairs, the interaction
potential between two electrons separated by a distance $r$ is given
by
\begin{equation}
V(r) = - \frac{G_{F}^{2}}{4\pi^{3}r^{5}},
\label{eq:vacpotm0}
\end{equation}
where $G_{F}$ is the Fermi constant, thus confirming the $1/r^5$ behaviour.
Hsu and Sikivie \cite{Hsu:1992tg} obtained the same result using Feynman
diagrammatic methods. 

The effect of the neutrino mass on the potential was first introduced in
Ref.~\cite{Fischbach:1996qf} for a single massive Dirac neutrino species, and in
Ref.~\cite{Grifols:1996fk} for the Majorana case; 
obtaining a potential with a finite range ${\cal O}(m_\nu^{-1})$.
The effects from several neutrino masses and the associated
flavor mixing was recently addressed in 
Refs.~\cite{Lusignoli:2010gw,LeThien:2019lxh,Segarra:2020rah,Costantino:2020bei}.  In Ref.~\cite{Bolton:2020xsm} the general form
of the neutrino forces generated by SM and beyond the Standard Model (BSM)
interactions in the framework of effective field theories was derived.
Finally, Ref.~\cite{Xu:2021daf} has included interactions beyond
four-fermion contact interactions.

To sum up, the existence of the 2-$\nu$ mediated long-range force
is a well-grounded prediction of the SM, but this force is
extremely weak. The $G_F^2$ suppression together with the $r^{-5}$ dependence
implies that it is already much weaker than gravity at distances
${\cal O}$(nm). This renders the effect orders of magnitude below present 
and near-future sensitivity of experiments testing the
gravitational inverse-square law~\cite{Hoskins:1985tn,Tan:2020vpf,Lee:2020zjt}
and the weak equivalence principle~\cite{Smith:1999cr,Schlamminger:2007ht}.
Notice, however, that refs.~\cite{Stadnik:2017yge,erratum}  pointed out that the very singular form of the potential may open up the posibility of improved sensitivity over $\sim$ fm distances  with atomic and nuclear spectroscopy.

However, if the interaction takes place in the presence of a background
of neutrinos, the 2-$\nu$
mediated potential could be significantly
modified. This was first discussed by Horowitz and Pantaleone
\cite{Horowitz:1993kw} and later by Ferrer {\sl et al.} ~\cite{Ferrer:1998ju,Ferrer:1999ad}, who evaluated the 2-$\nu$ mediated potential in the presence of 
the cosmic neutrino background (C$\nu$B). For this case they found a modest enhancement.
Most recently, the effect of neutrino backgrounds has been restudied and
extended in Ref.~\cite{Ghosh:2022nzo}. This work claims that the intense
neutrino fluxes from the Sun, supernovae (SN), or nuclear reactors,
may substantially enhance the potential (by up to 20 orders of
magnitude), in particular in the direction of the incoming background
neutrino flux. Such an enhancement could render the 2-$\nu$ potential
close to the sensitivity of current and near future fifth force
experiments,
which motivated the revision of the effect performed in the rest of
this note.

\section{Neutrino-mediated forces in a neutrino background: general remarks}

The key technical ingredient to evaluate
neutrino background effects in
Refs.~\cite{Horowitz:1993kw,Ferrer:1998ju,Ferrer:1999ad,Ghosh:2022nzo}
is the use of the background-modified propagator  obtained in finite
temperature field theory (TQFT) (for reviews of TQFT, see for example
Refs.~\cite{Kapusta:2006pm,Laine:2016hma}),
\begin{equation}
  S_F(p)=(\slash\!\!\!p +m)\left[
    \frac{i}{p^2-m^2+i\epsilon}-2\,\pi\,\delta(p^2-m^2)
    \left(\theta(p^0)f_\nu(\vec p)+\theta(-p^0)f_{\overline\nu}(\vec p)
    \right)
    \right]\, ,
  \label{eq:proptqft}
\end{equation}  
where $f_{\nu}(\vec p)$ and $f_{\overline \nu}(\vec p)$ are the
momentum distributions of neutrinos and antineutrinos, respectively,
normalized so that $\int \frac{d^3\vec p}{(2\pi)^3}
f_{\nu(\overline\nu)}(\vec p)$ gives their number density.
In what follows we show that such a modified propagator does not
\emph{always} correctly quantify the effect of the neutrino background
in the evaluation of the 2-$\nu$-exchange potential. This is true, in
particular, when the finite width of the wavepackets of the neutrinos
of the medium are considered.

Our first observation is that the derivation of the
background-modified fermion propagator in TQFT (\cref{eq:proptqft})
relies on the assumption that the fermions in the background are in
thermal equilibrium, while neither the present C$\nu$B, nor the
neutrino fluxes from the Sun, SN, or nuclear reactors are in thermal
equilibrium close to Earth.  Interestingly, one can still 
use QFT techniques to show that Eq.~\eqref{eq:proptqft} may be used
for the neutrino propagator in the presence of \emph{a class of}
backgrounds, see~\cite{Ghosh:2022nzo}.  This derivation, however,
implicitly assumes that the neutrino background is well-described by
an incoherent superposition of plane waves.  Neither the present
C$\nu$B, nor the neutrino fluxes from the Sun, SN, or nuclear reactors
can be well-described as plane waves at \emph{all distance
scales}. Even more, on general grounds, the Pauli exclusion principle
forbids a superposition of fermionic wavepackets {with equal
  momenta} over scales smaller than the size of the wavepacket. This
results on constraints on the number density and energy distribution
of a background of fermions {at those distance scales}.
  
We proceed now to evaluating the $2$-$\nu$-exchange in the presence of
realistic neutrino backgrounds, accounting for the finite width of the
wavepackets describing the background state.

\section{Neutrino-mediated forces in a neutrino background: formalism}

The static interaction potential plays a central role in
non-relativistic scattering in quantum mechanics as well as in
describing classical forces.  To obtain it given a relativistic QFT,
recall that for an interaction between two fermions $f_1$ and $f_2$
located at positions $\vec r_1$ and $\vec r_2$, the aforementioned
potential reads,~\cite{Itzykson:1980rh}
\begin{equation}
  V(\vec{r}\equiv \vec r_1-\vec r_2) = - \int d^3\vec{q} \,e^{i\vec{q}\cdot\vec{r}}\, T^{f_1f_2}_{\rm NR}(\vec{q})\;,
\label{eq:potential}
\end{equation}
where $T_{\rm NR}^{ff'}(\vec{q})$ can be obtained  from the fully relativistic
scattering amplitude starting with the ${\cal S}$ matrix for the transition
$\ket{I}\rightarrow \ket{F}$ computed in QFT in momentum space
\begin{equation}
  {\cal S}_{IF} = \delta_{IF} + i(2\pi)^4\delta^4(p_F-p_I) T_{IF}^{f_1f_2}\,,
\end{equation}
as
\begin{equation}
 T_{\rm NR}^{f_1f_2}(\vec{q}) \equiv \frac{T_{IF}^{f_1f_2}}{4\,m_f\,m_{f'}}\,,
\end{equation}  
with the four-momentum transfer between the fermions given by
$q\equiv(0,\vec q)\equiv p'_1-p_1\equiv p_2-p'_2$, and the helicities of $f_1$ and 
$f_2$ not changing in the process.

In the SM the relevant weak interaction Lagrangian
leading to the  2--$\nu$ mediated potential can be written
in the effective four-fermion interaction approximation as
\begin{equation}
\mathcal{L_I}=-\frac{G_F}{2\sqrt{2}}\bar{\psi}_{\nu_i}\gamma^\mu
(1-\gamma_5)\psi_{\nu_j}\bar{\psi}_{f}\gamma_\mu
(g^f_{V_{ij}}-g^f_{A_{ij}}\gamma_5)\psi_f\,,
\end{equation}
where $\nu_i$ are the neutrino fields with masses
$m_{\nu_i}$ and $f$ indicates the fermions in the external legs.
The effective couplings are
\begin{align} 
g_{V_{ij}}^e & = 2 U_{ei}U_{ej}-(1-4s_w^2)\delta_{ij}\, , \\
g_{V_{ij}}^p & = (1-4s_w^2)\delta_{ij}\, ,  \\
g_{V_{ij}}^n & =-\delta_{ij}\, ,\\
g_{A_{ij}}^e & = g_{A_{ij}}^n = -g_{A_{ij}}^p =- \delta_{ij}\,.
\end{align}
In what follows, for simplicity we neglect leptonic mixing and
 assume Dirac neutrinos. Mixing effects and Majorana neutrinos
do not affect our conclusions and  can be easily accounted for.

\begin{figure}[!htbp]
\centering
\includegraphics[width=0.35\textwidth]{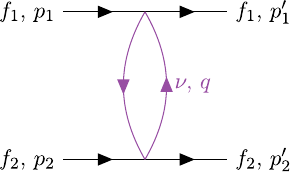}
\caption{Feynman diagram contributing to the relevant scattering amplitude
in vacuum}
\label{fig:fey0}
\end{figure}
The leading contribution to the 2-$\nu$ exchange potential
generated in vacuum can be obtained from the amplitude represented by the
Feynman diagram in Fig.~\ref{fig:fey0}. In the above notation, it 
corresponds to initial and final states 
$\ket{I} \equiv \ket{{p}_1,s_1}_{f_1}\ket{{p}_2,s_2}_{f_2}$
and $\ket{F} \equiv \ket{{p'}_1,s_1}_{f_1}\ket{{p'}_2,s_2}_{f_2}$, where
$p_i$ are the initial fermion 4-momenta (with $p^0_i=\sqrt{|\vec p_i|^2 + m_i^2}$) and 
$s_i$ their helicities. For massless neutrinos this leads
to the well-known result  
\begin{equation}
V(r)= -\frac{G_F^2\, g_V^{f_1}\, g_V^{f_2}}{16\,\pi^3} \frac{1}{r^5}  \;.
\end{equation}

\begin{figure}[!htbp]
\centering
 \includegraphics[width=0.35\textwidth]{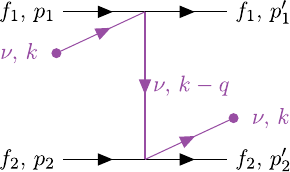} \hspace{0.1\textwidth} \includegraphics[width=0.35\textwidth]{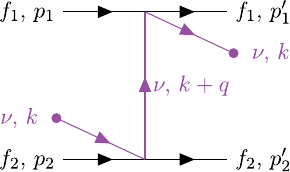} 
\caption{Feynman diagrams contributing to the relevant scattering amplitudes
 in the presence of a neutrino background.}
\label{fig:bkgFey}
\end{figure}

In the presence of a neutrino background new diagrams contribute to
the amplitude at the same order in perturbation theory. More
concretely, there is a process capturing a neutrino from the
background with momentum $\vec{k}$, exchanging a virtual neutrino, and
returning another neutrino with momentum $\vec{k}$ as represented in
Fig.~\ref{fig:bkgFey} (below, we justify that both background
neutrinos have the same momentum). The contributions from an
antineutrino background can be trivially obtained by inverting the
neutrino lines.\footnote{Notice that a diagram where both a neutrino and an antineutrino are absorbed or emitted would modify the fermion energy and hence it does not contribute to the static potential. This is also true for other scattering processes beyond those we consider in Fig.~\ref{fig:bkgFey}, and  we ignore them.}

Being of the
same order in perturbation theory, these contributions should be taken
into account for computing the potential in any environment
where neutrinos are present.  
As
mentioned above, the neutrino backgrounds that we shall consider
are far from being
in thermal equilibrium, and may even entail  momentum distributions  far from
thermal. 
This is why the use of TQFT techniques is not justified \emph{a priori}, and we will study the influence of the background explicitly. 

The basic assumption in what follows is that the background can be well
described by an incoherent superposition of neutrino
wavepackets of the form\footnote{One can add a random phase at each momentum state, but it does not play any role for the final observable and we ignore it.}
\begin{equation}
\ket{\psi(\vec{x},\vec{k}',s)} =
\int\frac{d^3\vec{k}}{(2\pi)^3\sqrt{2E_{k}}}
\omega(\vec{k}',\vec{k})
e^{-i\vec{k}\cdot\vec{x}}\ket{\vec{k},s}_\nu\,,
\end{equation}
where $\omega(\vec{k}',\vec{k})$ are neutrino wavepackets
centered at momentum of $\vec{k}'$ and with helicity $s$. The exponential factor
centers the packet at position $\vec{x}$.
In what follows, for concreteness in
our quantifications we use Gaussian wavepackets,
\begin{equation}
  |\omega(\vec{k}',\vec{k})|^2=
  \frac{1}{\pi^{3/2} \sigma^3}e^{-\frac{|\vec k '-\vec k|^2}{\sigma^2}}\,,
\label{eq:wp}    
\end{equation}
with $\sigma$ the width in momentum space (unlike the position-space width, that increases with time for free particles, this stays constant when the particles evolve freely).

To account for the stochastic properties of the medium, we parametrize
it in terms of a density matrix.  Assuming that the background is
homogeneous (this can be realized by large enough densities over the
probed distances or by averaging the force over a long enough time),
\begin{equation}
\begin{split}
\rho^{bkg} & = \sum_s \int d^3\vec{k}' f_\nu(\vec{k}')\int d^3\vec{x}
\ket{\psi(\vec{x},\vec{k}',s)}\bra{\psi(\vec{x},\vec{k}',s)}  \\
& =  \sum_s \int d^3\vec{k}' f_\nu(\vec{k}') \int\frac{d^3\vec{k}}{(2\pi)^3\sqrt{2E_{k}}} |\omega(\vec{k}',\vec{k})|^2 \ket{\vec{k},s}_\nu \bra{\vec{k},s}_\nu
\label{eq:rho}
\end{split}
\end{equation}
with $f_\nu(\vec{k}')$ the momentum distribution of the neutrino ensemble
which we assume to be spin-independent. For polarized
ultrarelativistic backgrounds, such as solar or reactor neutrinos,
only one of the helicities contributes, and hence this density matrix
is equally applicable. As we see, when computing averages both
background neutrinos will have the same momentum.

Before proceeding to the calculation of the potential, it is worth
pointing out that the Pauli exclusion principle imposes relevant
constraints on the possible form of a state composed of identical
fermions. For a thermal neutrino background for which
$f_\nu(\vec{k}')$ is a Dirac-Fermi distribution, Eq.~\eqref{eq:rho} is
fully consistent for any width of the wave packet.  On the contrary,
for a non-thermal neutrino background, characterized by a general
momentum distribution \emph{with arbitrary normalization}, the
previous description of the background as an incoherent sum of
\emph{single-particle states} is only physical for interparticle
distances larger that the spatial extent of the wavepackets.  For interparticle distances shorter than the extent of the
wavepacket, the Pauli exclusion principle renders the assumed momentum
distribution inconsistent.
{That said, in the physical scenarios we consider the background  densities
(ie the normalization of the $f_\nu$) are low enough, and  the interparticle
distances are always larger than wavepacket extents.
So in what follows we are quantifying exclusively the effects associated
with the inclusion of the wavepacket, under the assumption that the
normalization of the momentum distribution is consistent with the Pauli
exclusion principle over the distances studied.}

In this approach, we first compute the amplitudes in \cref{fig:bkgFey}
and we later trace out the result with the state that describes the background,
Eq.~\eqref{eq:rho}.
This corresponds to the following initial and final states in the
evaluation of the amplitude
\begin{align}
    \ket{I} &=
    \ket{p_1,s_1}_{f_1}\ket{p_2,s_2}_{f_2}\ket{k,s}_\nu \, ,
    \\ \ket{F} &=
    \ket{p'_1,s_1}_{f_1}\ket{p'_2,s_{2}}_{f_2}\ket{k,s}_\nu \, .
\end{align}
After some rearrangements, the corresponding ${\cal S}$ matrix element
can be written as,
\begin{equation}
\begin{split}
 {\cal  S}_{IF} = -\frac{G_F^2}{8}\int d^4x_1\,d^4x_2
  \;\null_{f_1}\!\!
  \bra{p'_1,s_1}\bar{\psi}_{f}({x}_1)\gamma_\mu
    (g^f_{V}-g^{f}_{A}\gamma_5)\psi_{f}({x}_1)
    \ket{p_1,s_1}_{f_1} \times
   \\ \null_{f_2}\!\!\bra{p'_2,s_2}
    \bar{\psi}_{f}({x}_2)\gamma_\nu
    (g^f_{V}-g^f_{A}\gamma_5)\psi_f({x}_2)
    \ket{p_2,s_2}_{f_2} \times
    \\ \Bigl[\null_\nu\!\!\bra{{k},s}
      \bar{\psi}_{\nu}({x}_1)\gamma^\mu
      (1-\gamma_5)\,i S_F(x_1-x_2)\,
      \gamma^\nu (1-\gamma_5)\psi_{\nu}({x}_2) \ket{k,s}_\nu +
      \\\null_\nu\!\! \bra{k,s} \bar{\psi}_{\nu}({x}_2)\gamma^\nu
      (1-\gamma_5)\, i S_F(x_1-x_2)\,\gamma^\mu
      (1-\gamma_5)\psi_{\nu}({x}_1) \ket{{k},s}_\nu \Bigr]\,,
\end{split}
\label{eq:sif}
\end{equation}
where $S_F(x_1-x_2)$ is the vacuum Feynman neutrino propagator in position space.

Fourier-expanding the fermion fields
in terms of creation and annihilation operators with well-determined momenta
and using the Fourier-space neutrino propagator,
one can easily compute the expectation values. After factorizing out
the global energy-momentum conservation Dirac delta, 
$i\,(2\pi)^4\delta^4(p_F-p_I)=i\,(2\pi)^4\delta^4(p_1+p_2-p'_1-p'_2)$, 
and taking the non-relativistic limit we get,
\begin{equation}
\begin{split}
\sum_s T^{f_1f_2\,\nu}_{NR} = - G_F^2g^{f_1}_Vg^{f_2}_V&\Biggl(\int d^4x\int
\frac{d^4q'}{(2\pi)^4}\frac{[2k^0q^0-\vec{k}\cdot\vec{q'}]}{q'^2-m^2}
e^{-ix(q-q'+k)} \\  &~~~~~+\int d^4x\int
\frac{d^4q'}{(2\pi)^4}\frac{[2k^0q^0-\vec{k}\cdot\vec{q'}]}{{q'^2-m^2}}
e^{-ix(q+q'-k)}\Biggr)\,,
\end{split}
\end{equation}
where the superindex $\nu$ indicates that the background neutrino state is still not traced out. 
Using \cref{eq:potential} and computing the trace with the
density matrix in \cref{eq:rho}, we arrive  at the following form
for the potential between the two fermions,
\footnote{Notice that the results only depend on the squared modulus
  of the momentum-space wavepacket$|\omega(k, k')|^2$.
  They are hence unaffected by position-space broadening induced by the
  past time evolution.}
\footnote{We notice in passing that this wavepacket-size effect on 
the {\sl background-mediated} potential, does not occur for
{\sl background-sourced} potentials, like the MSW potential for example,
because in such cases the potential probes the background particle fields in a
single point (while the background-mediated potential probes the
background particle fields at two points, see \eqref{eq:sif}).
Consequently if one evaluates the relevant matrix element 
using a density matrix ~\eqref{eq:rho} 
the result is independent of $\vec{k}$, the integral over
$d^3\vec{k}$ is immediate, and any dependence on $\omega$ disappears
from the expressions.}
\begin{equation}
\begin{split}
V^{\rm bkg}_\sigma(r)=& \, G_F^2g^{f_1}_Vg^{f_2}_V\int d^3\vec{q}\,e^{i\vec{q}\cdot\vec{r}}\int
\frac{d^3\vec{k}'}{(2\pi)^3}\,f_\nu(\vec{k}')\,\int
\frac{d^3\vec{k}}{(2\pi)^32E_k}\,|\omega(\vec{k}',\vec{k})|^2 \\
& \hspace*{4cm}\times \Biggl[
  \frac{2|\vec{k}|^2+m^2+\vec{k}\cdot\vec{q}}{|\vec{q}|^2+2\vec{k}\cdot\vec{q}}
  +
  \frac{2|\vec{k}|^2+m^2-\vec{k}\cdot\vec{q}}{|\vec{q}|^2-2\vec{k}\cdot\vec{q}}\Biggr]\,.\label{eq:potwave}
  \end{split}
\end{equation}

We find that the same result could have been obtained
starting with the amplitude corresponding to the diagram in Fig.~\ref{fig:fey0} 
and using an {\sl effective neutrino propagator in the neutrino medium}
given by
\begin{eqnarray}
  S_F(p)&=&(\slash\!\!\!p+m)\left\{
  \frac{i}{p^2-m^2+i\epsilon}\nonumber 
  -2\,\pi\,\delta(p^2-m^2)\right.\\
&&\hspace*{-2cm}\left.\times\left[\theta(p^0)    
    \int d^3{\vec{p}\,}'
    |\omega_\nu(\vec p,{\vec{p}\,}')|^2 f_\nu({\vec{p}\,}') +
    \theta(-p^0)
    \int d^3{\vec{p}\,}'
    |\omega_\nu(\vec p,{\vec{p}\,}')|^2 f_{\overline\nu}({\vec{p}\,}') 
    \right]\right\}\, \,,
  \label{eq:propour}
\end{eqnarray}  
which  reproduces
the thermal propagator \cref{eq:proptqft} in the $\sigma\rightarrow 0$ limit.

In fact, we can formally recover the results in the literature from
\cref{eq:potwave} by taking the limit of neutrino wavepackets to be
plane waves
\begin{equation} 
\lim_{\sigma\rightarrow 0}\, |\omega(\vec{k}',\vec{k})|^2=
\delta^3(\vec{k}'-\vec{k}) \;.
\label{eq:limdel}
\end{equation}
In particular, if $f(\vec{k}')$ is a thermal distribution we recover
the results obtained for the C$\nu$B~\cite{Horowitz:1993kw,Ferrer:1998ju,Ferrer:1999ad,Ghosh:2022nzo}.
Notice however, that as discussed above, the Pauli exclusion principle
makes this limit unphysical for the non-thermal neutrino backgrounds
from the Sun, nuclear reactors, or SN.

\section{Results and conclusions}

From the previous expressions, one can extract
the  corrections of the width of wavepackets in the 
2-$\nu$ exchange potential. 
Although the integrals in \cref{eq:potwave} cannot be performed analytically
for most $f_\nu(\vec p)$ distributions, the basic lessons can be extracted from 
the case of a monochromatic
directional flux $\Phi_0$
\begin{equation}
f_\nu(\vec p)=(2 \pi)^3 \delta^3(\vec p-\vec p_0) \Phi_0  \,,
\end{equation}
with $\vec{p}_0$ parallel to $\vec r$, which allows for an analytic
treatment. This example can be used to
estimate the modification of the potential due to solar, SN, and
reactor neutrinos.  Furthermore, \emph{this is the scenario for which the largest
background-induced enhancement is obtained  in Ref.~\cite{Ghosh:2022nzo}}.
For this case, and neglecting the neutrino mass,  we find for $E_\nu=|\vec k_0|$
\begin{align}
  V^{\rm bkg}_\sigma(r)&=\, -\frac{G_F^2\, g_V^{f_1}\,g_V^{f_2}}{4\ \pi^{3/2}\, r^2} \Phi_0
    \frac{1}{(4 \hat E^2 +\hat\sigma^4)^2}
    \left\{-2 e^{-\frac{\hat E_\nu^2}{\hat\sigma^2}} \hat\sigma
    (\hat\sigma^2-2 \hat E^2)(4 \hat E^2 +\hat\sigma^4)\right.\nonumber\\
    &
    +\sqrt{\pi}\left[2 \hat E\left(8 \hat E^4 +2 (1+\hat E)^2\hat\sigma^4+
      \hat\sigma^6\right){\rm Erf}\left[\frac{\hat E}{\hat\sigma}\right]\right.
      \nonumber \\
&\left.      + e^{-\hat\sigma^2}\left(8 \hat E^4+\hat\sigma^6+2\hat\sigma^8+
      \hat E^2(6\hat\sigma^4-4\hat\sigma^2)\right)
      \left(\cos(2\hat E){\rm Im}(X)+\sin(2\hat E){\rm Re}(X)\right)\right. \nonumber
      \\ &
      \left.\left.
      + e^{-\hat\sigma^2}
    \left(4 \hat E\hat\sigma^2(2 \hat E^2+\hat\sigma^2+\hat\sigma^4)\right)
    \left(\cos(2\hat E){\rm Im}(X)-\sin(2\hat E){\rm Re}(X)\right)
    \right]\right\}\;,
\label{eq:potgauss}    
\end{align}
where for convenience we have introduced the dimensionless variables
$\hat E \equiv E_\nu \, r$ and $\hat\sigma\equiv \sigma\, r$
(with $r=|\vec r|$), and
$X\equiv{\rm Erf}\left[\frac{\hat E^2}{\hat\sigma^2}+ i\,\hat\sigma\right]$.

\begin{figure}[hbtp]
\begin{subfigure}{0.49\textwidth}
    \centering \includegraphics[width=\textwidth]{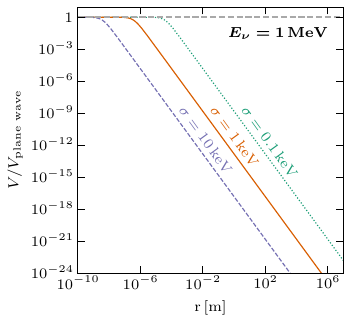}
\end{subfigure} \begin{subfigure}{0.49\textwidth}
    \centering \includegraphics[width=\textwidth]{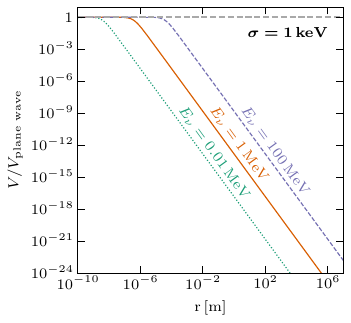}
\end{subfigure} 
    \caption{Suppression of the background-induced 2-$\nu$-mediated
      potential between two fermions due to the inclusion of a finite
      width packet of the background states as a function of the distance
      between the fermions. The figures illustrate the dependence of the
      effect on the neutrino energy and the width of the wavepacket.
    }
    \label{fig:results_1}
\end{figure}

The comparison of the potential in \cref{eq:potgauss} with the expression
obtained for $\sigma\rightarrow 0$
\begin{equation}
  V^{\rm bkg}_{\sigma\rightarrow0}(r)=-\frac{G_F^2\, g_V^{f_1}\,g_V^{f_2}}{4\pi\, r} \Phi_0
  \left[ E_\nu+\frac{\sin(2E_\nu r)}{2r}\right]\,,
\label{eq:potpw}    
\end{equation}
is shown in Fig.~\ref{fig:results_1}, where we plot 
$V^{\rm bkg}_\sigma(r)/V^{\rm bkg}_{\sigma\rightarrow0}(r)$ as a function
of $r$ for energies
around $1 \, \mathrm{MeV}$, characteristic of solar, SN, and reactor neutrinos;
and different wavepacket widths. As we see from the figure, including the
wavepacket width suppresses the effect of the background
by orders of magnitude for distances probed by the most sensitive
current experiments (which range from $\mu m$ to $\sim 10^4$ km). 

For widths $\sigma \ll E_\nu$ and distances $r \gg\frac{1}{E_\nu}$,
\cref{eq:potgauss} is well approximated by
\begin{equation}
    V^{\rm bkg}_\sigma(r) \simeq \frac{V^{\rm bkg}_{\sigma\rightarrow0}(r)}{1 + r^2 \sigma^4
      / (4 E_\nu^2)} \, .
\end{equation}
Although in \cref{fig:results_1} we use the full expression, this
approximation is excellent. From the figure one reads that to have an
unsuppressed potential due to a
background of neutrinos of energies ${\cal O}$(MeV) over a distance of
$\mu m$, the background neutrino states should have momentum widths
$\sigma\ll 10$~keV. For longer distances, $\sim 10^4 \, \mathrm{km}$,
the width has to be even smaller, $\sigma \ll \, \mathrm{meV}$.

Estimations on the typical wavepacket sizes for neutrino backgrounds
differ in the literature. For thermal systems with temperature $T$,
collisions act as quantum-mechanical measurements that localize the
particles, and $\sigma \sim T$~\cite{Akhmedov:2017mcc}. This would
lead to $\sigma \sim \mathrm{keV}$ for solar neutrinos (assuming that
neutrinos inherit the wavepacket width of their parent nuclei), and
$\sigma \sim \mathrm{MeV}$ for SN
neutrinos~\cite{Akhmedov:2017mcc}.
This implies that for a solar (SN)
neutrino of 1 MeV, the potential is suppressed over distances
larger than $\sim 10^{-6}$ m ($\sim 10^{-12}$ m). 
For reactor neutrinos, estimates are more uncertain, ranging
from $\sim 20\,\mathrm{MeV}$ to $\sim 1
\, \mathrm{eV}$~\cite{Akhmedov:2017mcc,deGouvea:2021uvg,Arguelles:2022bvt,Akhmedov:2022bjs,Jones:2022hme}, with corresponding suppression of the potential
from neutrinos of 1 MeV over distances larger than  $\sim 10^{-14}$ m
to $\sim$ m.

Overall, we find that the long-discussed 2-$\nu$ mediated force
remains undetectable for fifth-force experiments and the discussed
neutrino backgrounds. The effects only seem to be unsuppressed at
extremely short distances (c.f.~\cref{fig:results_1}), making it
difficult to envision physical scenarios able to probe
neutrino-induced long-range forces.

\textbf{Note added}: after our paper appeared on arXiv,
Ref.~\cite{Ghosh:2022nzo} updated their calculation to include
spectral smearing effects. We emphasize that the spectral
smearing they consider is physically different from the quantum-mechanical
wavepacket size effect we discuss. In particular, after
including spectral smearing they still find an enhancement in the
2-$\nu$ exchange potential for directional fluxes in directions
very close to the incoming flux (in our notation, $\vec{p}_0$
parallel to $\vec{r}$). Our results show that the wavepacket width
suppresses the potential irrespective of the direction.
Therefore, contrary to their statement, we do not agree with
their results.

\acknowledgments
We thank L. Badurina, O. Eboli, J. Ghiglieri, C. Manuel, N. Sabti, J. Taron and R. Zukanovich for valuable discussions. We thank Y. Grossman for comments on an earlier draft. This project has received funding/support from the European Union's Horizon
2020 research and innovation program under the Marie Skłodowska-Curie
grant agreement No 860881-HIDDeN, as well as from grants
PID2020-115845GBI00 and PID2019-105614GB-C21,
``Unit of Excellence Maria de Maeztu 2020-2023"
award to the ICC-UB CEX2019-000918-M,  
funded by MCIN/AEI/10.13039/501100011033.
MCGG is also supported by USA-NSF grant PHY-1915093.
DB is supported by a `Ayuda Beatriz Galindo Senior' from the Spanish
`Ministerio de Universidades', grant BG20/00228. IFAE is partially
funded by the CERCA program of the Generalitat de Catalunya.
\bibliographystyle{JHEP}
\bibliography{nupotential}
\end{document}